\documentclass[runningheads]{llncs}
\usepackage{graphicx}

\usepackage{amsmath}
\usepackage{amssymb}
\usepackage{color}
\usepackage[noadjust]{cite}
\usepackage{multirow}
\usepackage{tikz}
\usepackage{lineno}
\usepackage{algorithm2e}

\usepackage{subfig}
\usepackage{float}
\usepackage{hyperref}
\usepackage{xcolor}
\hypersetup{
    colorlinks=true,
    linkcolor=blue,
    urlcolor=blue,
    citecolor=blue
}
\usepackage{cleveref}

\newtheorem{observation}{Observation}

\nopagebreak

\date{}

\begin{document}

\title{Growth Rate of the Number of Empty Triangles in the Plane}

\author{Bhaswar B. Bhattacharya\inst{1}\and
Sandip Das\inst{2}\and
Sk Samim Islam\inst{2} \and
Saumya Sen\inst{2}}
\authorrunning{Bhattacharya, Das, Islam, and Sen }
%
\institute{University of Pennsylvania, USA \and
Indian Statistical Institute, Kolkata, India }
\maketitle              
\begin{abstract}
Given a set $P$ of $n$ points in the plane, in general position, denote by $N_\Delta(P)$ the number of empty triangles with vertices in $P$. In this paper we investigate by how much $N_\Delta(P)$ changes if a point $x$ is removed from $P$. 
By constructing a graph $G_P(x)$ based on the arrangement of the empty triangles incident on $x$, we transform this geometric problem to the problem of counting triangles in the graph $G_P(x)$. We study properties of the graph $G_P(x)$ and, in particular, show that it is kite-free. 
This relates the growth rate of the number of empty triangles to the famous Ruzsa-Szemer\'edi problem.  
\end{abstract}

\section{Introduction }\label{S:Intro}



Let $P$ be a set of $n$ points in the plane in general position, that is, no three are on a line. We define $N_{\triangle}(P)$ as the number of empty triangles in $P$, that is, the number of triangles with vertices in $P$ with no other point of $P$ in the interior. Counting the number of empty triangles in planar point sets is a classical problem in discrete geometry (see \cite{barany2004planar,barany1987empty,barany2013many,erdos1992geometry,garcia2011note,pinchasi2006empty,researchbook,reitzner2018stars,valtr1992} and the references therein). Specifically, B{\'a}r{\'a}ny and F{\"u}redi \cite{barany1987empty} showed that $N_\Delta(P) \geq  n^2-O(\log n)$, for any set of points $P$, with $|P|=n$, in general position. On the other hand, a set of $n$ points chosen uniformly and independently at random from a convex set of area 1 contains $2n^2 + o(n^2)$ empty triangles on expectation \cite{reitzner2018stars,valtr1992}.

\begin{figure}[ht]
     \centering
        \includegraphics[width=0.4\textwidth,height=0.4\textwidth]{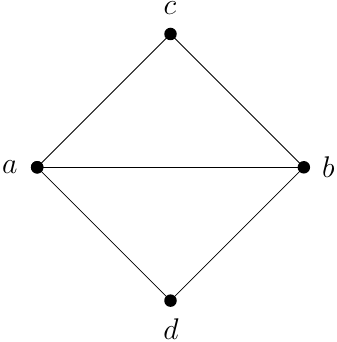}
         \caption{ The kite graph $K_4 \setminus \{ e \}$.}
         \label{fig:K4e}
    \end{figure}

In this paper we study the growth rate of $N_{\triangle}(P)$ when a point $x$ is removed from $P$.  For this, let $N_{\triangle}(P\setminus\{x\})$ denote the number of empty triangles in the set $P\setminus\{x\}$ and consider the difference:  
$$\Delta(x, P) = |N_{\triangle}(P)-N_{\triangle}(P\setminus\{x\})|.$$ 
In the following theorem we bound the above difference in terms of number of triangles in $P$ with $x$ as a vertex, which we denote by $V_P(x)$. To this end, denote by $K_4\setminus\{e\}$ the {\it kite graph}, that is, the complete graph $K_4$ with one of its diagonals removed (see Figure \ref{fig:K4e}). 



%
%

\begin{theorem}\label{triangle_containig_x} For any set $P$, with $|P| =n$, 
\begin{align}\label{eq:triangleVH}
\Delta(x, P) \leq V_P(x) + H(V_P(x), K_3, K_4\setminus\{e\}), 
\end{align}
where $H(V_P(x), K_3, K_4\setminus\{e\})$ is the maximum number of triangles in a $K_4\setminus\{e\}$-free graph on $V_P(x)$ vertices. Moreover, there exists a set $P$, with $|P|=n$, and a point $x \in P$ such that $\Delta(x, P) \geq C V_P(x)^{\frac{3}{2}}$, for some constant $C > 0$. 
\end{theorem} 

%
%

The proof of Theorem \ref{triangle_containig_x} is given in Section \ref{S:2a}. 
To establish the upper bound in \eqref{eq:triangleVH} we construct a graph $G_P(x)$ based on the arrangement of empty triangles in $P$ which have $x$ as a vertex and relate the problem of estimating $\Delta(x, P)$ to the problem of counting the number of triangles in the graph $G_P(x)$. The graph $G_P(x)$ has many interesting properties, specifically, we show that it is kite-free, which gives the upper bound in \eqref{eq:triangleVH} (see Section \ref{sec:upperboundpf}). The lower bound construction is given in Section \ref{sec:example}.

The extremal number $H(V_P(x), K_3, K_4\setminus\{e\})$ that appear in the bound \eqref{eq:triangleVH} is closely connected to the celebrated Ruzsa-Szemer\'edi problem, which asks for the maximum number of edges in a graph with $n$ vertices such that every edge belongs to a unique triangle \cite{ruzsa1978triple}. Note that if a graph $G$ is kite-free, then every edge in $G$ has at most one triangle passing through it. Then removing the edges which have no triangles passing through them (which does not change the number of triangles), one can relate the problem of counting triangles in $G$ to the Ruzsa-Szemer\'edi problem. An application of the Szemer\'edi regularity lemma \cite{szemeredi1975regular} shows that any solution to the  Ruzsa-Szemer\'edi problem has at most $o(n^2)$ edges \cite{ruzsa1978triple} (which is often also  referred to as the diamond-free lemma). This can be improved to $n^2 /e^{\Omega(\log^{\star} n)}$ by a stronger form of the graph removal lemma \cite{graph2011}. 
This, in particular, implies that $H(V_P(x), K_3, K_4\setminus\{e\}) = o(V_P(x)^2)$. Hence, from Theorem \ref{triangle_containig_x} we get the upper bound:  $\Delta(x, P) = o(V_P(x)^2)$. 
On the other hand, the lower bound of $\Delta(x, P)$ from Theorem \ref{triangle_containig_x} is  $\Omega (V_P(x)^{\frac{3}{2}})$. We believe the correct order of magnitude of $\Delta(x, P)$ is closer to the lower bound, because the graph $G_P(x)$ has additional geometric structure. We collect some geometric properties of the graph $G_P(x)$ in Section \ref{sec:graphproperties}, which we believe can be of independent interest.

\section{Proof of Theorem \ref{triangle_containig_x}} 
\label{S:2a} 

We prove the upper bound in \eqref{eq:triangleVH} in Section \ref{sec:upperboundpf}. The lower bound construction is given in Section \ref{sec:example}. 

\subsection{Proof of the Upper Bound} 
\label{sec:upperboundpf} 

We begin with the following simple observation: 

   \begin{observation}\label{n_1+n_2}
        $\Delta(x, P)\leq V_P(x) + I_P(x)$, 
    where $I_P(x)$ is the number of triangles in $P$ that contain only the point $x$ in the interior. 
    \end{observation} 
    
    \begin{proof} 
    Let $U_P(x)$ denote the number of empty triangles in $P$ such that $x$ is not a vertex of the empty triangles. Note that $N_{\triangle}(P)=V_P(x)+U_P(x)$ and 
            $N_{\triangle}(P\setminus\{x\})=U_P(x) + I_P(x)$. This implies, $|N_{\triangle}(P)-N_{\triangle}(P\setminus\{x\})|
            = |V_P(x) - I_P(x)| \leq V_P(x) + I_P(x)$. \hfill $\Box$
    \end{proof} 
    


\begin{figure}[ht]
            \centering
            \includegraphics[width=\textwidth]{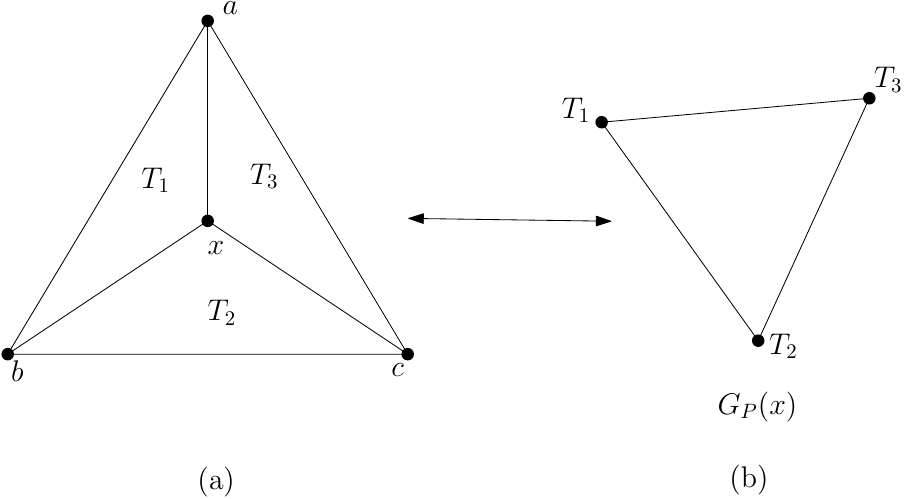}
            \caption{A set of points $P=\{x, a, b, c\}$ and the empty triangle graph incident at $x$. }
            \label{figure Geometry to Graph transformation}
        \end{figure}

         
        Given a set $P$, with $|P| =n$, and a point $x \in P$, define the graph $G_P(x)$ as follows: The vertex set of $G_P(x)$ is $V(G_P(x))$, the set of triangles in $P$ with $x$ as one of their vertices, and there should be edge between 2 vertices in $G_P(x)$ if the corresponding triangles, say $T_1$ and $T_2$, satisfy the following conditions: 
         \begin{itemize}
	\item $T_1$ and $T_2$ share an edge, 
	\item $T_1$ and $T_2$ are area disjoint, 
	\item the sum of angles of $T_1$ and $T_2$ incident at $x$ is greater than  $180^{\circ}$. 
         \end{itemize} 
         We call the graph $G_P(x)$ the {\it empty triangle graph incident at $x$}. 
         Figure \ref{figure Geometry to Graph transformation} shows the graph $G_P (x)$ for a set of 4 points $P=\{x, a, b, c\}$. (It is worth noting that we use $\Delta$ in notations that count empty triangles in the point set $P$ and $K_3$ to denote a triangle in the graph $G_P(x)$.) 
%
%

The following lemma shows that $I_P(x)$, as defined in Observation \ref{n_1+n_2}, is equal to the number of triangles in $G_P(x)$.

 \begin{lemma}\label{counting_N_22} Suppose $P$ be a set of points in the plane, with $|P| =n$, in general position and $x \in P$. Then 
$$I_P(x) = N_{K_3}(G_P(x)) , $$ 
where $N_{K_3}(G_P(x))$ is the number of triangles in the graph $G_P(x)$. 
  \end{lemma}
  
 \begin{proof}

 Let $(a, b, c)$ be a triangle in $I_P(x)$, that is, $(a, b, c)$ only has the point $x$ in the interior. This corresponds to a triangle in $G_P(x)$ as shown in Figure \ref{figure Geometry to Graph transformation}. 
 
 
 Now, consider a triangle with vertices labeled $(1, 2, 3)$ in $G_P(x)$. The vertices $1$, $2$, and $3$ in $G_P(x)$ correspond to 3 empty triangles in $P$ that have the point $x$ as one of their vertices, which we denote by $T_1$, $T_2$, and $T_3$, respectively. Since there is an edge between 1 and 2 in $G_P(x)$,  the triangles $T_1$ and $T_2$ in $P$ share a common edge, are area disjoint, and  the sum of angles of $T_1$ and $T_2$ incident at $x$ is greater than  $180^{\circ}$. Hence, we can assume, without loss of generality, $T_1$ and $T_2$ are arranged as in Figure \ref{fig:two_sub_figures}(a).  Similarly, the pairs of triangles $(T_2, T_3)$ and $(T_1, T_3)$ share a common edge, are area disjoint, and the sum of their angles incident at $x$ is greater than  $180^{\circ}$. This implies, $T_1$, $T_2$, and $T_3$ are mutually area disjoint, and they cannot all share the same edge. 
 Therefore, the triangles $T_1$, $T_2$, and $T_3$ have to be arranged as in Figure \ref{fig:two_sub_figures}(b). Note that the triangle formed by the union of the 3 triangles $T_1, T_2$ and $T_3$ only contains the point $x$ in $P$. Hence, for every triangle $(1, 2, 3)$ in $G_P(x)$  one gets a triangle in $I_P(x)$, formed by the union of the 3 triangles $T_1, T_2$ and $T_3$.  \hfill $\Box$

   \begin{figure}[ht]
    \centering
    \subfloat[\label{fig:subima}]{%
    \includegraphics[width=.25\linewidth]{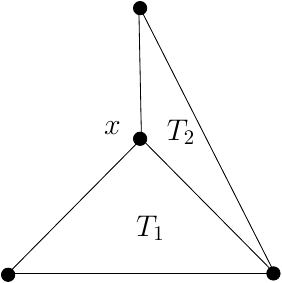}%
    } 
    \hfil
    \subfloat[\label{fig:subimb}]{%
    \includegraphics[width=.25\linewidth]{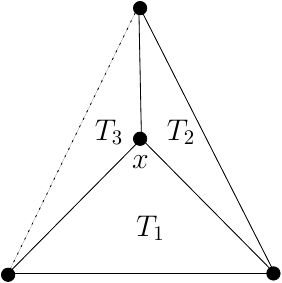}%
    }
    \caption{Illustration for the proof of Lemma \ref{counting_N_22}. }
    \label{fig:two_sub_figures}
  \end{figure}
    \end{proof}
     

 
Applying Observation \ref{n_1+n_2} and Lemma \ref{counting_N_22} it follows that 
\begin{align}\label{eq:triangleVHpf}
\Delta(x, P)\leq V_P(x) + I_P(x) = V_P(x) + N_{K_3}(G_P(x)). 
\end{align}
Therefore, to prove the upper bound in \eqref{eq:triangleVH}, it remains to show that 
\begin{align}\label{eq:Gtriangle}
N_{K_3}(G_P(x)) \leq H(V_P(x), K_3, K_4 \setminus \{e\})
\end{align}
This follows from the lemma below which shows that the graph $G_P(x)$ is kite-free.

\begin{lemma}\label{counting_N_2}
  The graph $G_P(x)$ does not contain $K_4 \setminus \{e\}$ as a subgraph, that is, $G_P(x)$ is kite-free.
\end{lemma} 

\begin{proof} 
  Suppose $G_P(x)$ contains a kite $K_4 \setminus \{e\}$, with vertices labeled $a, b, c, d$ as in Figure \ref{fig:K4e}. This corresponds to empty triangles $T_a, T_b, T_c, T_d$ with $x$ as a vertex such that $(T_a, T_b, T_c)$ and $(T_a, T_b, T_d)$ are mutually interior disjoint, and the pairs of triangles $(T_a, T_b)$, $(T_b, T_c)$, $(T_a, T_c)$, $(T_a, T_d)$, and $(T_b, T_d)$ share a common edge. This means the triangles $T_a, T_b, T_c$ must  be arranged as in Figure \ref{figure Geometry to Graph transformation}(a). Hence, it is impossible to place $T_d$ which share an edge with $T_a$ and $T_b$ and is interior disjoint from $T_a, T_b$, unless $T_d$ coincides with $T_c$. This shows $G_P(x)$ cannot contain a $K_4 \setminus \{e\}$ as a subgraph.  \hfill $\Box$ 
%
%
\end{proof}

\begin{figure}[ht]
     \centering      
         \includegraphics[width=0.68\textwidth]{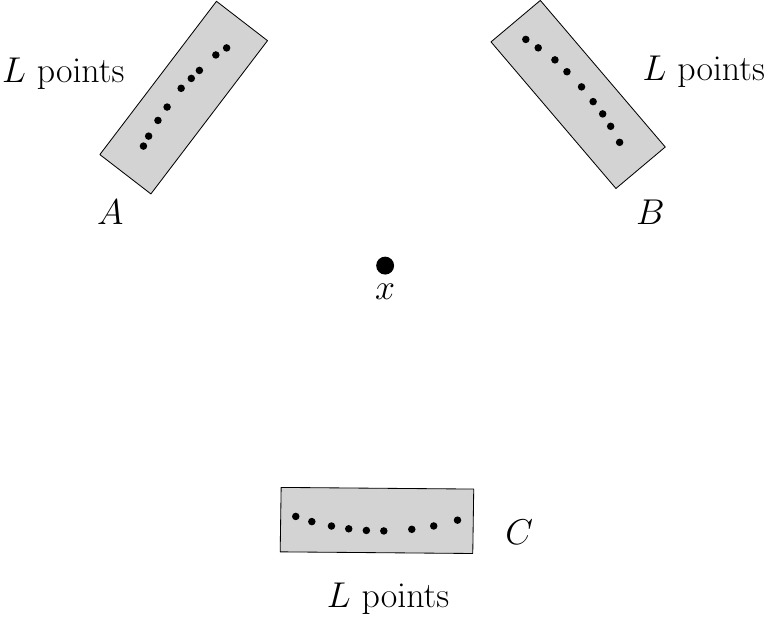}
         \caption{Example showing the lower bound in Theorem \ref{triangle_containig_x}. }  
         \label{fig:vertextriangle}
\end{figure}

Since $G_P(x)$ is kite-free by Lemma \ref{counting_N_2}, the bound in  \eqref{eq:Gtriangle} follows. This together with \eqref{eq:triangleVHpf} gives the upper bound in \eqref{eq:triangleVH}.  

\subsection{Lower Bound Construction} 
\label{sec:example}

To prove the lower bound in Theorem \ref{triangle_containig_x} consider the set of points $P$, with $|P|=n=3L+1$, as shown in Figure \ref{fig:vertextriangle}. Specifically, $P$ consists of three sets points $A$, $B$, and $C$, with $|A|= |B| = |C| = L$, arranged along 3 disjoint convex chains and a point $x$ at the middle.  
Note that $N_\Delta(P \setminus \{x\}) = {3L \choose 3} \sim \frac{9}{2} L^3$.  Also, 
\begin{align}\label{eq:VP}
V_P(x) = 3 {L \choose 2} + 3 L^2 = \Theta(L^2). 
\end{align} 
To compute $N_\Delta(P)$ recall that $N_\Delta(P) = V_P(x) + U_P(x)$, where $U_P(x)$ is as defined in Observation \ref{n_1+n_2}. Now, note that 
$$U_P(x) = 3 {L \choose 3} + 6 L {L \choose 2} \sim 3.5 L^3.$$
Hence, $N_\Delta(P) \sim 3.5 L^3 + \Theta(L^2)$ and $$\Delta(x, P) = |N_\Delta(P) - N_\Delta(P \setminus \{x\}) | = \Theta(L^3) = \Theta (V_P(x)^{\frac{3}{2}}), $$
from \eqref{eq:VP}. This completes the proof of the lower bound in Theorem \ref{triangle_containig_x}. 

%
%

\section{Properties of the Graph $G_P(x)$}
\label{sec:graphproperties}


In this section we collect some geometric properties of the graph $G_P(x)$. First, we show that $G_P(x)$ can contain arbitrarily large bipartite graphs.

\begin{lemma}\label{lm:bipartite} Fix $r, s \geq 1$. Then there exists a set of points $P$ and $x \in P$ such that the graph $G_P(x)$ contains the complete bipartite graph $K_{r, s}$. 
\end{lemma}

\begin{figure}[ht]
            \centering           
            \includegraphics[width=0.85\textwidth,height=0.4\textwidth]{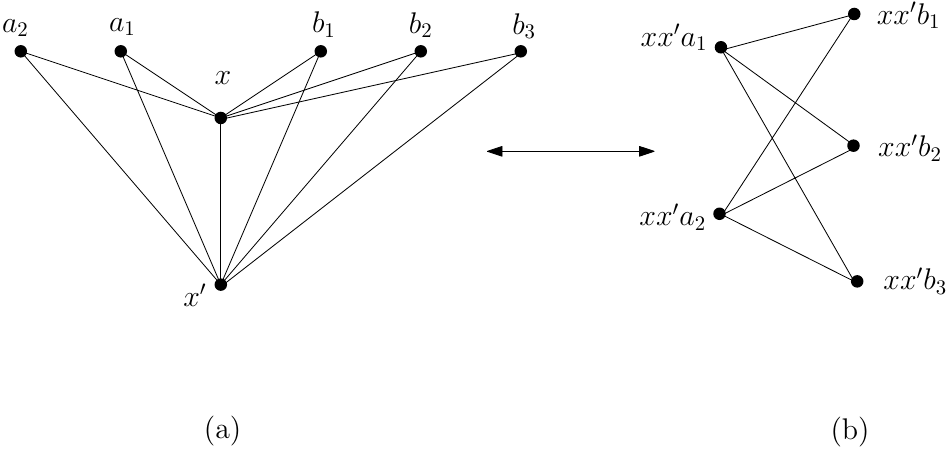}
            \caption{Illustration for the proof of Lemma \ref{lm:bipartite}.}
            \label{Bipartite Graph}
\end{figure}

\begin{proof} Consider the set of $r+s+2$ points $$P =\{x, x', a_1, a_2, \ldots, a_r, b_1, b_2, \ldots, b_s\} ,$$ as shown in Figure \ref{Bipartite Graph}(a) (with $r=2$ and $s=3$). Specifically, the points $\{a_1, a_2, $ $\ldots, a_r\}$ and $\{b_1, b_2, $ $\ldots, b_s\}$ lie on 2 disjoint convex chains and the points $x$ and $x'$ are in the middle. Note that, for all $1 \leq i \leq r$ and $1 \leq j \leq s$, the triangles $xx'a_i$ and $xx'b_j$ share a common edge, are interior disjoint, and the sum of the angles incident at $x$ is greater than $180^{\circ}$. This implies that the graph $G_P(x)$ contains the complete bipartite graph $K_{r, s}$.   \hfill $\Box$ 
\end{proof} 

The reason it is worthwhile to know whether or not $\Delta(x, P)$ contains complete bipartite graphs as subgraphs, is because of a possible approach to improve the $o(n^2)$ upper bound on $\Delta(x, P)$ through the K\H{o}v\'ari-S\'os-Tur\'an theorem \cite{zarankiewicz}. Recall that the K\H{o}v\'ari-S\'os-Tur\'an theorem states that any graph which is $K_{r, s}$-free, where $r \leq s$, has at most $O(n^{2-\frac{1}{r}})$ edges. Therefore, if $\Delta(x, P)$ did not contain some complete bipartite  graph as a subgraph, then it would have led to a polynomial improvement over the $o(n^2)$ upper bound on $\Delta(x, P)$. Lemma \ref{lm:bipartite} shows that this is not the case, hence, one cannot directly apply the  K\H{o}v\'ari-S\'os-Tur\'an theorem to improve the upper bound on $\Delta(x, P)$.

Although the K\H{o}v\'ari-S\'os-Tur\'an result cannot be directly applied to the graph $G_P(x)$, we believe a polynomial improvement over the Ruzsa-Szemer\'edi upper bound on $\Delta_P(x)$ is possible, because the graph $G_P(x)$ has additional geometric structure. To illustrate this we show that the well-known Behrend's construction \cite{behrend}, which gives a nearly quadratic lower bound on Ruzsa-Szemer\'edi problem, is not geometric realizable. We begin recalling Behrend's construction:
%
%
%
%
%


\begin{definition}[Behrend's graph] \label{defn:geometric}
Suppose $p$ is an odd prime and $A \subseteq \mathbb Z/p \mathbb Z$ is a set with no 3-term arithmetic progression. The Behrend's graph $G(p, A)$ is a tripartite graph with vertices on each side of the tripartition numbered $\{0, 1, \ldots, p-1\}$ and triangles of the form $(z, z+a, z+2a)$ modulo $p$, for $z \in \{0, 1, \ldots, p-1\}$ and $a \in A$. 
\end{definition} 

\begin{figure}[ht]
     \centering
        \includegraphics[width=0.65\textwidth]{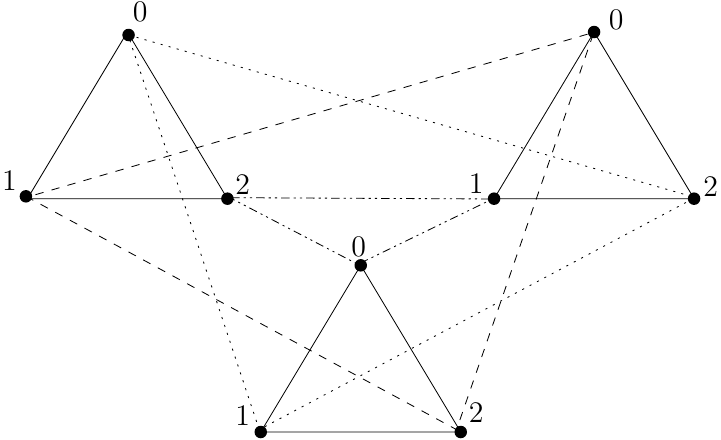}
         \caption{The Paley graph with 9 vertices, 18 edges, and 6 triangles. }
         \label{figure:vertex9}
\end{figure}

It is easy to check that the graph $G(p, A)$ has $3p$ vertices $3|A|p$ edges and each edge belongs to a unique triangle. For example, when $p=3$ and $A =\{1, 2\}$ one gets the 9 vertex Paley graph shown in Figure \ref{figure:vertex9}. Behrend \cite{behrend} constructed a set $A$ of size $p/e^{O(\sqrt{\log p})}$ with no 3-term arithmetic progression. Using this set in the Behrend's graph in Definition \ref{defn:geometric} one gets a lower bound of $\Omega (p^2/e^{O(\sqrt{\log p})})$ for the Ruzsa-Szemer\'edi problem and, hence, for $H(n, K_3, K_4 \setminus \{e\})$.

%
%

The following result shows that the graph in Figure \ref{figure:vertex9} cannot be geometrically realized, that is, it is not possible to find a set of points $P$ and $x \in P$ such that $G_P(x)$ is isomorphic to the graph in Figure \ref{figure:vertex9}.

\begin{proposition}\label{ppn:geometric}
The graph in Figure \ref{figure:vertex9} is not geometrically realizable.  
\end{proposition}

Proposition \ref{ppn:geometric} shows that Behrend's graphs are not geometrically realizable. This, in particular, illustrates that the graph $G_P(x)$ has a richer geometric structure than the collection of kite-free graphs. 

\subsection{Proof of Proposition \ref{ppn:geometric}}  
\label{sec:geometricpf}

 We proceed by contradiction. Suppose there exists a point set $P$ and $x \in P$ such that $G_P(x)$ is isomorphic to the graph in Figure \ref{figure:vertex9}. This implies $V_P(x)=9$ and $I_P(x)= N_{K_3}(G_P(x))= 6$ (by Lemma \ref{counting_N_22}). This, in particular,  means that there are 6 triangles in $P$ which only contains the point $x$ in the interior. Denote these triangles by  $\mathcal T=\{T_1', T_2', \ldots, T_6' \}$. 

%
%

    \begin{figure}[ht]
     \centering
             \includegraphics[width=0.45\textwidth]{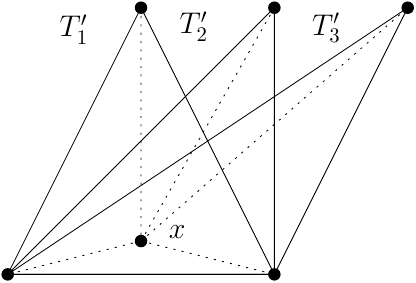}%
            \caption{Illustration for the proofs of Lemma \ref{lm:triangleedge}. }
     \label{fig:image23}
    \end{figure}     
    
   \begin{lemma}\label{lm:triangleedge}
   There cannot be 3 triangles in the set $\mathcal T$ which share a common edge. 
   \end{lemma}  
   
\begin{figure}[ht]
  \centering
         \subfloat[\label{fig:subim2}]{%
         \includegraphics[width=0.45\textwidth]{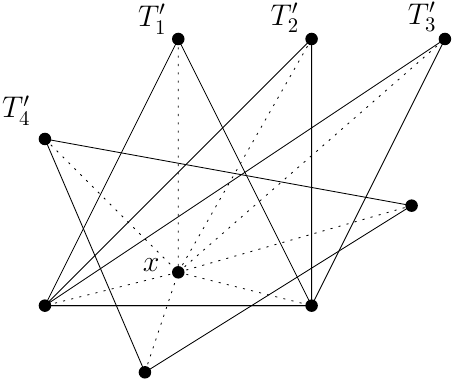}%
         }\hfil
         \subfloat[\label{fig:subim2}]{%
         \includegraphics[width=0.45\textwidth]{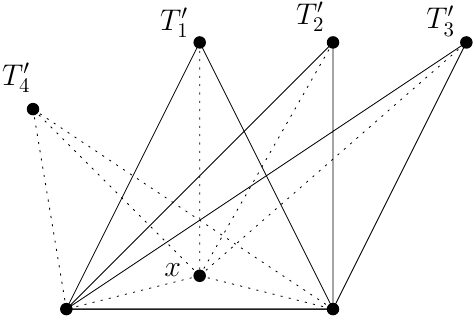}%
         }
    \caption{Illustrations for the proofs of Lemma \ref{lm:triangleedge}: (a) Case 1 and (b) Case 2. }
     \label{fig:image2}
\end{figure}

   \begin{proof} Let, if possible, there exists 3 triangles $T_1', T_2', T_3'$ in $\mathcal T$ which share one edge (see Figure \ref{fig:image23}(a)). 
 As all the triangles contain $x$ as the only interior point they must have a common  shareable area that contains $x$. Therefore, it is not possible to create a new triangle here by joining vertices of existing triangles which will contain $x$ . Therefore, to create a new triangle which will contain $x$ as the only interior point, the following two cases may occur. 

\begin{itemize}  

\item[$\bullet$] {\it Case} 1: There is a triangle $T_4' \in \mathcal T$ that does not share any edge and vertex with $T_1', T_2', T_3'$ (see Figure \ref{fig:image2} (a)) or there is a triangle $T_4' \in \mathcal T$ that does not share any edge with $T_1', T_2', T_3'$ but shares one vertex with $T_1', T_2', T_3'$. In both case $V_P(x) \geq 10$, which is a contradiction. 


\item[$\bullet$] {\it Case} 2: There is a triangle $T_4' \in \mathcal T$ which shares an edge with the triangles $T_1', T_2', T_3'$. This means $T_4'$ has a vertex that is not common with the vertices of $T_1', T_2', T_3'$ (see Figure \ref{fig:image2} (b)) In this case $V_P(x) = 9$ and $I_P(x)= N_{K_3}(G_P(x)) = 4$. Consider the triangle $T_5'$ in $\mathcal{T}\setminus \{T_1', T_2', T_3',T_4'\}$. Note that, there is at least one edge in $T_5'$ which does not belong to the edge set of the triangles $T_1', T_2', T_3',T_4'$. This edge forms an empty triangle whose one vertex is $x$. This implies  $V_P(x) \geq 10$, which is a contradiction. 

\end{itemize}

Thus, if there are 3 triangles in the set $\mathcal T$ which share an edge, then it is not possible to place another 3 triangles satisfying the required geometric constraints for the graph in Figure \ref{figure:vertex9}. Hence, there cannot be 3 triangle in $\mathcal T$ which share an edge. 
\hfill $\Box$
   \end{proof} 

Observe that each edge of a triangle in $\mathcal T$ generates an empty triangle with vertex $x$. 
Since there are 9 empty triangles with vertex $x$ in $G_P(x)$, the number of distinct edges generated by the triangles in $\mathcal T$ will be 9. Moreover, by Lemma  \ref{lm:triangleedge} no edge of a triangle in $\mathcal T$ can be shared by three triangles of $\mathcal T$. Hence, each of the 9 edges generated by the triangles in $\mathcal T$ must belong to exactly 2 triangles in $\mathcal T$ (since, counting with repetitions, there are a total of $6 \times 3 =18$ edges in the triangles in $\mathcal T$). 
%
%
The following lemma shows that this is not geometrically realizable.

\begin{lemma}\label{lm:triangle3G} 
All $3$ edges of any triangle in $\mathcal T$ cannot be shared  by other triangles in $\mathcal T$. 
\end{lemma}

\begin{figure}[ht]
     \centering
         \subfloat[\label{fig:subim1}]{%
             \includegraphics[width=0.40\textwidth]{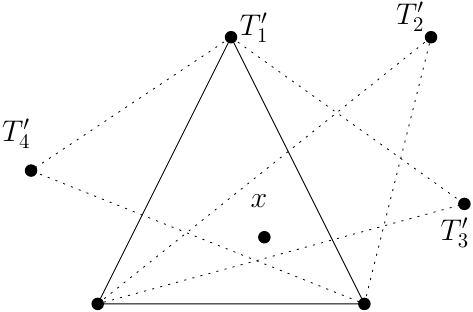}%
         }\hfil
         \subfloat[\label{fig:subim2}]{%
         \includegraphics[width=0.50\textwidth]{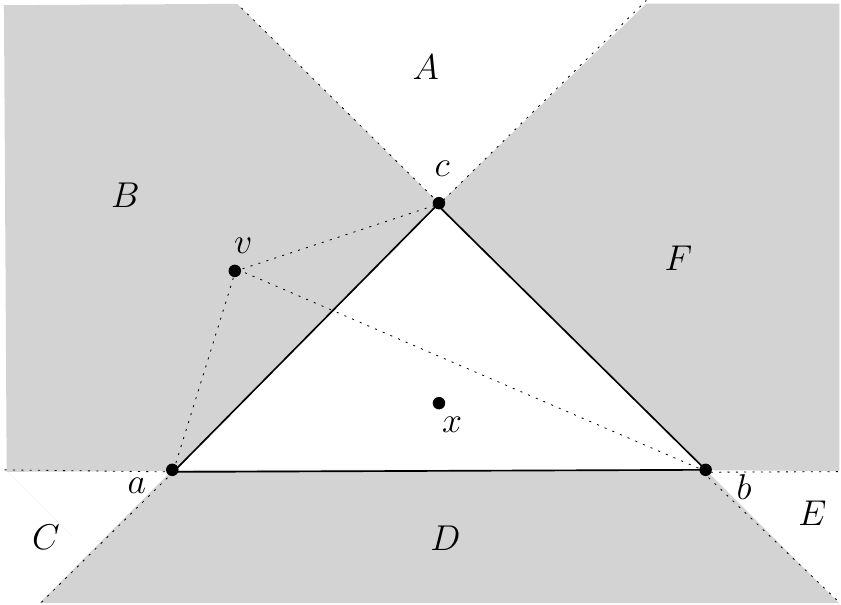}%
         }
    \caption{Illustrations for the proofs of  Lemma \ref{lm:triangle3G}. }
     \label{fig:image3}
\end{figure}

 \begin{proof} 
 Let, if possible, there exists a triangle $T_1' \in \mathcal T$ whose all $3$ edges are shared by the triangles $T_2', T_3', T_4'$ and $T_2', T_3', T_4'$ do not share edges in between. (see Figure \ref{fig:image3} (a)). In this case  $V_P(x)=9$ and $I_P(x)= N_{K_3}(G_P(x))= 4$, which is impossible by arguments similar to Case 2 of Lemma \ref{lm:triangleedge}. 
 
 Alternatively, suppose there exists a triangle $T_1' = (a, b, c) \in \mathcal T$, such that the edge $(a, b)$ is shared by the triangle $T_2'$, the edge $(b, c)$ is shared by the triangle $T_3'$, the edge $(c, a)$ is shared by the edge $T_4'$, and $T_2'$ and $T_3'$ share a common edge.  This implies, there is a vertex $v$ such that $T_2'=(a, b, v) $ and $T_3'= (b, c, v)$ (see Figure \ref{fig:image3} (b)). Note that $v$ cannot be inside the triangle $(a, b, c)$, since $(a, b, c)$ has only $x$ as the interior point. Also, $v$ cannot be in region $A$, region $C$, and region $E$, because then either the triangle $T_2'$ or the triangle $T_3'$ will have more than one point in the interior. Hence, $v$ has to be in region $B$, region $D$, or region $F$. 
%
%
In this case, the triangle $T_2'$ and the triangle $T_3'$ are area disjoint, hence only one of them can contain $x$ in the interior (see Figure \ref{fig:image3} (b)). This gives a contradiction and completes the proof of the lemma. 
 \hfill $\Box$ 
 \end{proof} 

%
%

\noindent
Combining Lemma \ref{lm:triangleedge} and Lemma \ref{lm:triangle3G}, the result in Proposition \ref{ppn:geometric} follows.

\section{Conclusions}\label{sec conclusions}

In this paper, we initiate the study of the growth rate of the number of empty triangles in the plane, by proving upper and lower bounds on the difference $\Delta(x, P)$. We relate the upper bound to the well-known  Ruzsa-Szemer\'edi problem and study geometric properties of the triangle incidence graph $G_P(x)$. Our results show that $\Delta(x, P)$ can range from $O(V_P(x)^{\frac{3}{2}})$ and $o(V_P(x)^2)$. Understanding additional properties of the graph $G_P(x)$ is an interesting future direction, which can be useful in improving the bounds on $\Delta(x, P)$.


\bibliography{ref.bib}

\end{document}